\documentclass{article}
\usepackage{spconf,amsmath,graphicx}
\usepackage{amssymb}
\usepackage{booktabs}
\usepackage{multirow}
\usepackage{bbding}
\usepackage{hyperref}
\usepackage[numbers,sort]{natbib}
\title{VSANet: Real-time speech enhancement based on voice activity detection and causal spatial attention}
\name{Yuewei Zhang$^{1\ast}$ \qquad Huanbin Zou$^{2\ast}$ \qquad Jie Zhu$^{1\dagger}$\thanks{$^{\ast}$ Equally contribute to this work.} \thanks{$^{\dagger}$ Corresponding author (Email: zhujie@sjtu.edu.cn).}}
\address{$^{1}$ Department of Electronic Engineering, Shanghai Jiao Tong University, Shanghai, China \\
${^2}$ Tencent Video Cloud, Shanghai, China}
\copyrightnotice{979-8-3503-0689-7/23/\$31.00~\copyright2023 IEEE}
\begin{document}
%
\maketitle
\begin{abstract}
The deep learning-based speech enhancement (SE) methods always take the clean speech’s waveform or time-frequency spectrum feature as the learning target, and train the deep neural network (DNN) by reducing the error loss between the DNN’s output and the target. This is a conventional single-task learning paradigm, which has been proven to be effective, but we find that the multi-task learning framework can improve SE performance. Specifically, we design a framework containing a SE module and a voice activity detection (VAD) module, both of which share the same encoder, and the whole network is optimized by the weighted loss of the two modules. Moreover, we design a causal spatial attention (CSA) block to promote the representation capability of DNN. Combining the \textbf{V}AD aided multi-task learning framework and C\textbf{SA} block, our SE \textbf{net}work is named \textbf{VSANet}. The experimental results prove the benefits of multi-task learning and the CSA block, which give VSANet an excellent SE performance.
\end{abstract}
\begin{keywords}
speech enhancement, voice activity detection, multi-task learning, causal spatial attention block
\end{keywords}
\section{Introduction}
\label{sec:intro}
Speech enhancement (SE) is a fundamental task in speech signal processing, whose purpose is suppressing environmental noise, thereby improving the perceptual quality and intelligibility of speech. SE has been widely applied in various fields, such as speech communication, automatic speech recognition (ASR), and hearing assistant devices. Traditional SE methods \cite{1163209,1164550} always depend on specific prior assumptions and manual experience. However, their performance tends to be unsatisfactory in scenarios with highly non-stationary noise and low signal-to-noise (SNR).

Recently, the deep neural network (DNN) has achieved great success in the SE task. DNN-based SE algorithms can be divided into two categories: the time domain methods \cite{pascual2017segan,defossez20_interspeech} and the time-frequency (TF) domain methods \cite{tan18_interspeech,hu20g_interspeech,9642159}, with the latter being more commonly used.
The TF domain methods first convert the one-dimensional (1D) noisy speech into a two-dimensional (2D) noisy TF spectrum using a TF transformation technique. The DNN is then employed to process the 2D noisy spectrum and perform noise suppression, resulting in an enhanced spectrum. Finally, the enhanced spectrum is converted back to 1D enhanced speech using the corresponding inverse TF transformation method. TF domain methods can be further categorized into mapping methods \cite{8682834,8910352} and masking methods \cite{hu20g_interspeech,pmlr-v97-fu19b}. Mapping methods directly estimate the enhanced spectrum, while masking methods estimate the spectrum mask. In this paper, we adopt the masking method for SE and we utilize the convolutional recurrent network (CRN) \cite{tan18_interspeech} as the backbone of our SE network, which follows an encoder-decoder (ED) structure. Meanwhile, we employ short-time discrete cosine transform (STDCT) \cite{1672377} instead of the commonly used short-time Fourier transform (STFT) for TF transformation \cite{li2021real,9689230}. This choice avoids the phase estimation problem associated with STFT-based methods and reduces the model complexity.

Meanwhile, conventional SE approaches typically follow a single-task optimization paradigm, where the sole objective of the DNN is to predict the enhanced speech. However, many studies have demonstrated that a multi-task learning scheme achieves superior performance compared to single-task learning \cite{1716818,4721436}. The idea behind multi-task learning is to share the representation extracted by a public network to jointly optimize several related tasks. The knowledge learned from one subtask can assist in the learning of other subtasks. Multi-task learning has been widely used in speech signal processing \cite{10022924}, computer vision \cite{7423818}, and natural language processing \cite{7918389}. In the context of SE, some previous works have also explored similar strategies. \cite{Zheng_Peng_Zhang_Srinivasan_Lu_2021} proposed simultaneous estimation of speech and noise spectrum, which were then used to predict the spectrum mask. And \cite{app10093230} introduced a similar idea. In \cite{9616112}, the authors took the speech presence probability (SPP) as an auxiliary task to improve the SE performance. 

In this paper, we propose a multi-task learning framework that combines the voice activity detection (VAD) task with the SE task. As an important front-end processing module for many applications such as ASR \cite{5740583} and speaker verification \cite{9003935}, VAD aims to distinguish speech segments from noise segments. Prior works \cite{wang15e_interspeech,9414445} have already attempted to combine the SE task and VAD task to construct a multi-task learning framework. However, these works predominantly employed the SE network as a preprocessing or auxiliary module to improve VAD accuracy. In contrast, we believe that VAD can also benefit SE. Specifically, correct labeling of speech and noise segments can help the SE network to completely suppress the signal of noise segments. Besides, it also enables the SE network to focus on the feature learning of speech segments, facilitating better suppression of additive noise and better preservation of speech components in speech segments. Therefore, we consider VAD as the secondary task in our multi-task learning framework to assist the learning of SE task. Meanwhile, the two tasks' networks share the same encoder, and the loss function of the overall framework is a weighted sum of the two tasks’ losses. Experimental results demonstrate that incorporating the VAD task further improves the performance of the original SE network.

Besides, the attention mechanism \cite{Woo_2018_ECCV} has been proven to significantly improve the performance of various DNN-based tasks. Inspired by the attention mechanism of human perception, researchers have developed different types of attention modules. These attention modules guide the DNN to focus on important features while disregarding unnecessary information, thereby boosting the feature extraction, representation, and modeling capabilities of DNN. In our CRN-based SE model, both the encoder and decoder process 2D TF spectrum features. However, the useful feature information of speech is not uniformly distributed across the 2D TF feature maps. Thus, we propose to use a lightweight spatial attention block to assist in high-level feature extraction in the encoder and feature reconstruction in the decoder. In addition, we strictly guarantee the causality of our spatial attention block, ensuring the causality and real-time processing of our SE model. Our ablation study confirms the benefits of our designed causal spatial attention (CSA) block to the SE model’s performance.

Our contributions can be summarized as follows:
\begin{itemize}
    \item We adopt STDCT for TF transformation, resulting in a real-valued spectrum. This approach avoids the phase estimation problem in STFT-based methods, which can be difficult and computationally expensive.
    
    \item We integrate the SE task with the VAD task to create a multi-task learning framework. In this framework, the VAD network is used to aid the learning of the SE network, so as to further improve the SE performance.

    \item We design a CSA block, which improves the performance of the CRN-based SE model while ensuring the causality of the model.
\end{itemize}

Since we introduce a \textbf{V}AD aided multi-task learning framework with C\textbf{SA} block to achieve SE, the overall \textbf{net}work is abbreviated as \textbf{VSANet}. 


\section{Method}
\label{sec:method}
\subsection{Signal Model with Discrete Cosine Transform}
The purpose of SE is to estimate the clean speech $s(n)$ ($n$ is the discrete time index) from the noisy speech $x(n)$, which is polluted by additive noise $z(n)$ as
\begin{equation}
x(n)=s(n)+z(n)
\label{equation1}
\end{equation}

In our chosen TF domain method for SE, the first step is to transform the noisy speech signal $x(n)$ into the TF spectrum. To achieve this, we utilize the STDCT, a real-valued TF transformation that contains implicit phase information. The definition of discrete cosine transform (DCT) \cite{1672377} is
\begin{equation}
X(\mu)=c(\mu) \sqrt{\frac{2}{N}} \sum_{n=0}^{N-1} x(n) \cos \left[\frac{\pi \mu(2 n+1)}{2 N}\right]
\label{equation2}
\end{equation}
and the corresponding inverse DCT is
\begin{equation}
x(n)=\sqrt{\frac{2}{N}} \sum_{\mu=0}^{N-1} c(\mu) X(\mu) \cos \left[\frac{\pi \mu(2 n+1)}{2 N}\right]
\label{equation3}
\end{equation}
where $\mu\in{[0, 1, ..., N-1]}$ represents $N$ frequency bins from low to high, $n\in{[0, 1, ..., N-1]}$ represents $N$ different time indexes, and $c(\mu)$ in Eq. (\ref{equation2}) and Eq. (\ref{equation3}) is defined as
\begin{equation}
c(\mu)=\left\{\begin{array}{rlrl}
\frac{1}{\sqrt{2}}, & \mu=0 \\
1, & \mu=1,2, \ldots, N-1
\end{array}\right.
\label{equation4}
\end{equation}
 
Considering the short-time stationary characteristics of speech, we always divide it into overlapping short-time frames. These speech frames are then processed using a specific window function. Finally, the DCT is applied to each frame. This process allows us to obtain the STDCT spectrum of speech.

Thus, the TF domain form of Eq. (\ref{equation1}) can be written as
\begin{equation}
X=S+Z
\label{equation5}
\end{equation}
where the $X$, $S$ and $Z$ are the STDCT spectrum of $x$, $s$ and $z$, respectively.

\begin{figure*}[htbp]
\centerline{\includegraphics[height=3.1in]{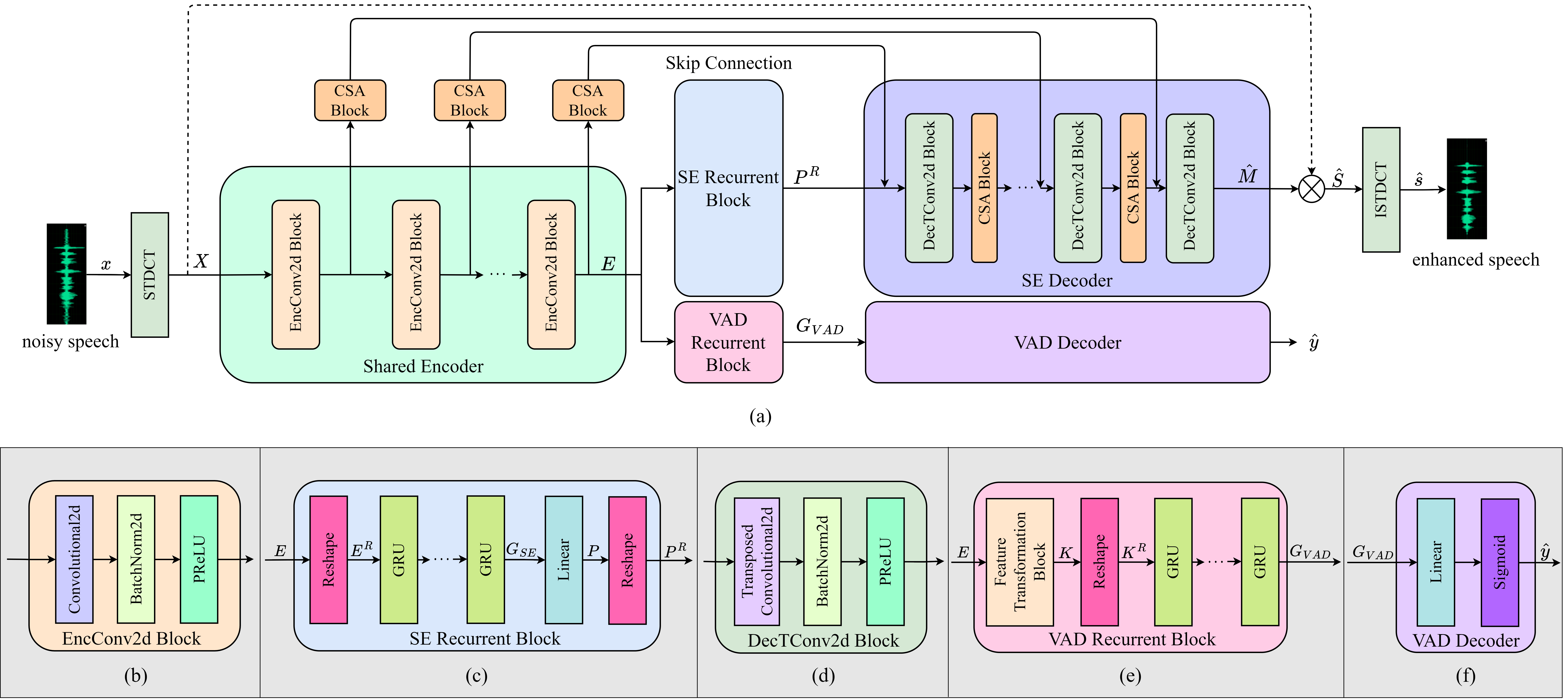}}
\caption{(a) Overall architecture of VSANet. (b) The detail of EncConv2d block. (c) The detail of speech enhancement (SE) recurrent block. (d) The detail of DecTConv2d block. (e) The detail of voice activity detection (VAD) recurrent block. (f) The detail of VAD decoder.} \vspace{-0.2cm} \label{fig1}
\end{figure*}

\subsection{Multi-task Learning Network Architecture}
The overall architecture of our multi-task learning framework is depicted in Fig. \ref{fig1}(a), comprising a SE module and a VAD module. In multi-task learning, parameter sharing achieves a regularization among highly correlated subtasks, leading to better generalization and performance \cite{10.1093/nsr/nwx105}. However, excessive parameter sharing among subtasks may hinder the learning process of individual subtasks. Therefore, we adopt a strategy where the SE and VAD modules only share the same encoder, while the remaining network structure is designed as a parallel dual-branch topology. In the following sections, we will introduce the details of the two subtasks, including the network structure, calculation process, and learning target.

\subsubsection{Speech Enhancement Module}
Our SE module utilizes a CRN-based structure. As shown in Fig. \ref{fig1}(a), it consists of a shared encoder, a SE recurrent block, and a SE decoder. The input noisy speech $x$ is initially transformed into the TF spectrum $X\in{\mathbb{R}^{1\times{F}\times{T}}}$, where $F$ and $T$ represent the frequency and frame dimensions, respectively. The TF spectrum $X$ is fed into the shared encoder, which includes multiple 2D convolutional (EncConv2d) blocks. The detail of the EncConv2d block is illustrated in Fig. \ref{fig1}(b). Each EncConv2d block contains a 2D convolutional layer, a 2D batch normalization layer, and a PReLU layer. The purpose of the shared encoder is to extract high-level features from the noisy spectrum, resulting in the feature map $E\in{\mathbb{R}^{C'\times{F'}\times{T}}}$. Similar to the previous works \cite{hu20g_interspeech}, we ensure a constant time resolution of the feature information throughout the module's computation. Then, a SE recurrent block, consisting mainly of multiple gated recurrent unit (GRU) layers, receives the feature map from the encoder and models the temporal correlation in speech features. Fig. \ref{fig1}(c) illustrates the structure of the SE recurrent block. Specifically, the encoded feature map $E$ is reshaped to $E^{R}\in{\mathbb{R}^{T\times{C'F'}}}$ so that the GRU layers can capture the inter-frame context from $E^{R}$. In addition, to ensure the processed result of the SE recurrent block has the same shape as the input $E$, the GRU layers' output, denoted as $G_{SE}\in{\mathbb{R}^{T\times{H}}}$, is further mapped to $P\in{\mathbb{R}^{T\times{H'}}}$ using a linear layer. Here, we set $H'=C'F'$, allowing us to reshape the output $P$ to $P^{R}\in{\mathbb{R}^{C'\times{F'}\times{T}}}$, which aligns with the shape of $E$. Next, a SE decoder is designed to reconstruct the low-resolution features $P^{R}$ back to the original input spectrum size. As shown in Fig. \ref{fig1}(a), the SE decoder is composed of several 2D transposed convolutional (DecTConv2d) blocks and CSA blocks. The principles of the CSA block will be introduced in detail in section \ref{subsec:CSA}, and the structure of the DecTConv2d block is illustrated in Fig \ref{fig1}(d). Each DecTConv2d block comprises a 2D transposed convolutional layer, a 2D batch normalization layer, and a PReLU layer. Note that the non-linear activation function of the last DecTConv2d block is adjusted to a modified Tanh function like \cite{9182513}. Finally, the output of the SE decoder, denoted as $\hat{M}\in{\mathbb{R}^{1\times{F}\times{T}}}$, is an estimation for the learning target $M$ of the SE module. While the learning target $M$ corresponds to the DCT ideal ratio mask (DCTIRM) and can be derived as
\begin{equation}
M = \frac{S}{X}
\label{equation6}
\end{equation}

As a result, the enhanced STDCT spectrum is
\begin{equation}
\hat{S}=\hat{M}\cdot{X}
\label{equation7}
\end{equation}
and the enhanced speech $\hat{s}$ is
\begin{equation}
\hat{s}=\mathrm{ISTDCT}(\hat{S})
\label{equation8}
\end{equation}
where $\mathrm{ISTDCT}(\cdot)$ represents the inverse STDCT operation.

\begin{figure*}[htbp]
\centerline{\includegraphics[height=1.19in]{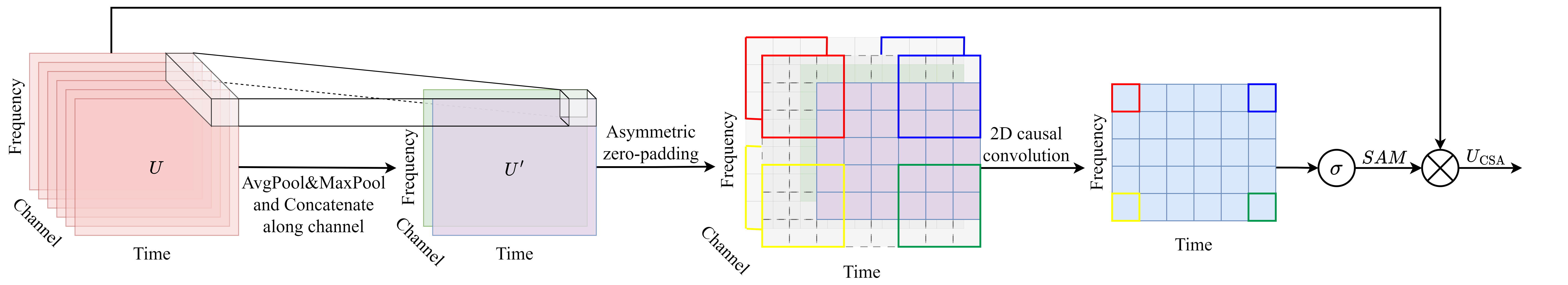}}
\caption{Causal spatial attention (CSA) block. For the calculation process of the 2D causal convolution, the bold boxes with four different colours identify four different inputs and their corresponding outputs during this process.} \vspace{-0.2cm} \label{fig2}
\end{figure*}

\subsubsection{Voice Activity Detection Module}
In contrast to the symmetric ED architecture of the SE module, the VAD module adopts an asymmetric setting. As illustrated in Fig. \ref{fig1}(a), the VAD module shares the encoder with the SE module and also includes a VAD recurrent block and a VAD decoder. Fig. \ref{fig1}(e) depicts the detail of the VAD recurrent block. Specifically, the output $E$ from the shared encoder is fed into a feature transformation block, which has the same structure as the EncConv2d block used in the shared encoder. Similar to the SE task, the transformed feature map $K\in{\mathbb{R}^{C''\times{F''}\times{T}}}$ is reshaped to $K^{R}\in{\mathbb{R}^{T\times{C''F''}}}$ and subsequently processed by several GRU layers. Although the GRU layers of the VAD module are not shared with the SE module, they have the same purpose of capturing the temporal dependencies in speech. Next, the output $G_{VAD}\in{\mathbb{R}^{T\times{H''}}}$ from the VAD recurrent block is fed into the VAD decoder. As shown in Fig. \ref{fig1}(f), the VAD decoder consists of a linear layer and a Sigmoid activation function layer. The linear layer maps the feature vector $G_{VAD}[t, :]\in{\mathbb{R}^{1\times{H''}}}$ at each frame $t$ to a value $\hat{v}_t\in{\mathbb{R}^{1}}$, and the Sigmoid function $\sigma$ further limits the range of $\hat{v}_t$ to $(0, 1)$ as
\begin{equation}
\hat{y}_t=\sigma(\hat{v}_t)
\label{equation9}
\end{equation}

Therefore, the VAD module finally generates the probability-like soft decision scores $\hat{\boldsymbol{y}}=[\hat{y}_0, \hat{y}_1, ..., \hat{y}_{T-1}]\in{(0, 1)^{T}}$ at the frame level. Correspondingly, the learning target of the VAD module is the ground-truth VAD label $\boldsymbol{y}=[y_0, y_1, ..., y_{T-1}]$, where each element $y_t\in{\{0, 1\}}$.

\subsection{Causal Spatial Attention}
\label{subsec:CSA}
We adopt the spatial attention mechanism \cite{Woo_2018_ECCV} to guide the network to focus on important regions of the feature map. Meanwhile, to guarantee the network’s causal configuration for real-time SE, we introduce a causal spatial attention (CSA) block. The CSA block exclusively relies on the current and past feature information to generate the attention map for the current frame.

The CSA block is illustrated in Fig. \ref{fig2}. To compute the spatial attention weights, we first apply 2D average-pooling and max-pooling operations along the channel dimension to the input feature map $U\in{\mathbb{R}^{C_i\times{F_i}\times{T_i}}}$. The two pooled single-channel feature maps are denoted as $U_{\text{avg}}\in{\mathbb{R}^{1\times{F_i}\times{T_i}}}$ and $U_{\text{max}}\in{\mathbb{R}^{1\times{F_i}\times{T_i}}}$, respectively. We then concatenate $U_{\text{avg}}$ and $U_{\text{max}}$ to obtain a dual-channel feature map $U'\in{\mathbb{R}^{2\times{F_i}\times{T_i}}}$. In previous works, a 2D convolutional layer with a kernel size of $k_F\times{k_T}$ is typically applied to $U'$, generating a spatial attention map $SAM\in{\mathbb{R}^{1\times{F_a}\times{T_a}}}$. To ensure that $SAM$ has the same width and height as $U$(i.e., $F_i=F_a$, $T_i=T_a$), symmetric zero-padding is commonly employed on $U'$ before the convolution operation. However, symmetric zero-padding introduces non-causality. Thus, we propose to use an asymmetric zero-padding operation in the time dimension of $U'$. Specifically, when padding zeros in the time dimension of $U'$, we only pad $(k_{T}-1)$ zeros on its starting side. As the zero-padding in the frequency dimension does not affect causality, it still employs symmetric zero-padding. In a word, the computation of the $SAM$ can be expressed as
\begin{equation}
U'=[\text{AvgPool}(U);\text{MaxPool}(U)]
\label{equation10}
\end{equation}
\begin{equation}
SAM=\sigma(f^{k_F\times{k_T}}(\text{AsymPad}(U')))
\label{equation11}
\end{equation}
where $[;]$ represents concatenation along the channel dimension, $\sigma$ is the Sigmoid function, and $f^{k_F\times{k_T}}(\cdot)$ denotes the convolution operation with an output channel of 1 and a kernel size of $k_F\times{k_T}$.

Eventually, the input feature map $U$ is multiplied with $SAM$ to obtain the final refined feature map as
\begin{equation}
U_{\text{CSA}}=SAM\otimes{U}
\label{equation12}
\end{equation}
where $\otimes$ represents the element-wise multiplication. During the multiplication operation, the attention weight is broadcasted along the channel dimension.

As shown in Fig \ref{fig1}(a), we incorporate the CSA block into the SE decoder and the skip connection paths to improve the feature representation capability of the SE module.

\subsection{Joint Loss Function}
In our multi-task learning framework, the total loss function $\mathcal{L}$ contains both SE loss $\mathcal{L}_{\mathrm{SE}}$ and VAD loss $\mathcal{L}_{\mathrm{VAD}}$. $\mathcal{L}$ is the weighted sum of the two loss components as
\begin{equation}
\mathcal{L} = \lambda_{1}\mathcal{L}_{\mathrm{SE}}+\lambda_{2}\mathcal{L}_{\mathrm{VAD}}
\label{equation13}
\end{equation}
where $\lambda_{1}$ and $\lambda_{2}$ are the weights for the SE loss and VAD loss, respectively.

The loss function $\mathcal{L}_{\mathrm{SE}}$ of the SE module is a joint loss as
\begin{equation}
\mathcal{L}_{\mathrm{SE}}=\mathcal{L}_{T}(s, \hat{s})+\alpha\mathcal{L}_{TF}(M, \hat{M})
\label{equation14}
\end{equation}
where $\mathcal{L}_{T}(s, \hat{s})$ represents the L1 norm loss between the clean speech $s$ and the enhanced speech $\hat{s}$, and $\mathcal{L}_{TF}(M, \hat{M})$ denotes the mean square error (MSE) loss between the target DCTIRM $M$ and the estimated mask $\hat{M}$. $\alpha$ is the weight.

As for the $\mathcal{L}_{\mathrm{VAD}}$ of the VAD module, it is the binary cross entropy (BCE) loss between the ground-truth VAD label $\boldsymbol{y}$ and the probability-like soft decision score $\hat{\boldsymbol{y}}$, i.e.,
\begin{equation}
\mathcal{L}_{\mathrm{VAD}}=\mathcal{L}_{\mathrm{BCE}}(\boldsymbol{y}, \hat{\boldsymbol{y}})
\label{equation15}
\end{equation}

In this paper, we set the above weights as $\lambda_{1}=1$, $\lambda_{2}=0.1$, and $\alpha=1$.

\section{Experiments}
\label{sec:exp}
\subsection{Dataset}
All experiments are conducted on the publicly available VoiceBank+DEMAND dataset \cite{valentinibotinhao16_interspeech}. The clean speech in this dataset is derived from the VoiceBank corpus \cite{6709856}, where 28 speakers’ recordings are used for training and the other 2 speakers’ recordings are used for testing. The training set consists of 11,572 clean-noisy speech pairs. Each clean speech is mixed with one of 10 noise types, including 2 artificially generated noises and 8 real noise recordings from the Demand database \cite{10.1121/1.4799597}. The mixed SNRs in the training set are \{0dB,5dB,10dB,15dB\}. The test set contains 824 speech pairs. The noisy utterances are generated by mixing the clean speech with 5 unseen noise types from the Demand database. The mixed SNRs in the test set are \{2.5dB,7.5dB,12.5dB,17.5dB\}. All utterances are resampled to 16 kHz.

\subsection{Experimental Setup}
In the experiments, we employ the Hamming window for STDCT. The window length and hop size are 32ms and 8ms, respectively. We use a 512-point STDCT, resulting in a frequency dimension of 512 for the STDCT spectrum.

In our VSANet, all the convolutional and transposed convolutional layers of the shared encoder, SE decoder, and the feature transformation block of the VAD module have the same setting. The kernel size and stride in frequency and time dimensions are set as (5,2) and (2,1), respectively. To ensure causality in the time dimension, asymmetric zero-padding is applied. The shared encoder consists of five EncConv2d blocks, and the output channel of each convolutional layer in the shared encoder is \{16,32,64,128,256\}. The SE recurrent block contains three GRU layers with hidden nodes of \{128,64,32\}. Following the GRU layers, there is a $32\times4096$ linear layer. The SE decoder is symmetric to the share encoder and consists of five DecTConv2d blocks. The output channel of each transposed convolutional layers is \{128,64,32,16,1\}. For the CSA block, the value of $k_F$ and $k_T$ are set as 7 and 15, respectively. In the VAD module, the output channel of the convolutional layer in the feature transformation block is 8. The hidden nodes of the VAD recurrent block’s GRU layers is \{32,16,8\}. In addition, the linear layer in the VAD decoder is set as $8\times1$. 

All the models are trained by RMSprop optimizer. The initial learning rate is 2e-4 and it decays by 0.5 if the performance does not improve for 6 consecutive epochs. The batch size and the total training epochs are 16 and 80, respectively.

\begin{table}[h]
  \centering
  \vspace{-0.2cm}
  \caption{Ablation study on VoiceBank+DEMAND test set}
  \setlength{\tabcolsep}{1mm}{
    \begin{tabular}{lccccc}
    \toprule[2pt]
          & WB-PESQ & CSIG  & CBAK  & COVL \\
    \hline
    noisy & 1.97  & 3.35  & 2.44  & 2.63 \\
    DCTCRN \cite{li2021real} & 2.85  & 4.14  & 3.45  & 3.52 \\
    DCTCRN+VAD & 2.92  & 4.18  & 3.49  & 3.55 \\
    DCTCRN+CSA & 2.91  & 4.16  & 3.49  & 3.55 \\
    DCTCRN+VAD+CSA & \multirow{2}*{\textbf{2.98}} & \multirow{2}*{\textbf{4.21}} & \multirow{2}*{\textbf{3.51}} & \multirow{2}*{\textbf{3.60}}\\
(i.e.,VSANet) & ~ & ~ & ~ & ~ & ~\\
    \bottomrule[2pt]
    \end{tabular}}%
    \vspace{-0.4cm}
  \label{table1}%
\end{table}%

\begin{table*}[ht]
  \centering
  \caption{Performance comparison with other state-of-the-art systems on VoiceBank+DEMAND test set under causal implementation. "-" indicates the result is not provided in the related paper.}
    \setlength{\tabcolsep}{2.9mm}{
    \begin{tabular}{lccccccccc}
    \toprule[2pt]
    Model    & Year  & Input & Cau.  & Param. (M) & WB-PESQ & CSIG  & CBAK  & COVL \\
    \hline
    noisy & -     & -     & -     & -     & 1.97  & 3.35  & 2.44  & 2.63 \\
    RNNoise \cite{8547084} & 2018 & Magnitude & \CheckmarkBold & 0.06 & 2.29 & - & - & - \\
    CRN \cite{tan18_interspeech}   & 2018  & Magnitude & \CheckmarkBold   & -  & 2.56  & 3.51  & 2.98  & 3.02 \\
    DCCRN \cite{hu20g_interspeech} & 2020 & Complex & \CheckmarkBold & 3.7 & 2.68 & 3.88 & 3.18 & 3.27 \\
    PercepNet \cite{valin20_interspeech} & 2020 & Magnitude & \CheckmarkBold & 8 & 2.73 & - & - & - \\
    DeepMMSE \cite{9066933} & 2020  & Magnitude & \CheckmarkBold  & -     & 2.77   & 4.14  & 3.32  & 3.46 \\
    DCCRN+ \cite{lv2021dccrn} & 2021  & Complex &  \CheckmarkBold  & 3.3   & 2.84    & -     & -     & - \\
    DEMUCS \cite{defossez20_interspeech} & 2021  & Time  & \CheckmarkBold  & 128   & 2.93   & 4.22  & 3.25  & 3.52 \\
    FullSubNet+ \cite{9747888} & 2022 & Complex\&Magnitude & \CheckmarkBold & 8.67 & 2.88 & 3.86 & 3.42 & 3.57 \\
    LFSFNet \cite{chen22c_interspeech} & 2022  & Magnitude & \CheckmarkBold  & 3.1   & 2.91    & -     & -     & - \\
    GaGNet \cite{LI2022108499} & 2022  & Complex & \CheckmarkBold  & 5.94  & 2.94   & \textbf{4.26} & 3.45  & 3.59 \\
    VSANet (ours) & 2023  & STDCT spectrum & \CheckmarkBold  & 3.1  & \textbf{2.98} & 4.21  & \textbf{3.51}  & \textbf{3.60} \\
    \bottomrule[2pt]
    \end{tabular}}%
    \vspace{-0.2cm}
  \label{tabel2}%
\end{table*}%

\subsection{Ablation Study}
To demonstrate the benefits of the VAD aided multi-task learning scheme and the CSA block to SE performance, we compare our proposed VASNet with the baseline model called DCTCRN \cite{li2021real}. DCTCRN uses STDCT for TF transformation and adopts the CRN structure as the network backbone. In order to control the experimental variables and ensure the reliability of the subsequent ablation study, the encoder, recurrent block, and decoder of DCTCRN are exactly the same as the SE module in our VSANet. The learning target of DCTCRN is also the DCTIRM as defined in Eq. (\ref{equation6}), and the enhanced speech is also reconstructed by Eq. (\ref{equation7}) and Eq. (\ref{equation8}). Moreover, DCTCRN employs the joint loss $\mathcal{L}_{\mathrm{SE}}$ as its loss function. Following the same training strategy as VSANet, we train DCTCRN on the VoiceBank+DEMAND dataset and evaluate its performance using four objective metrics, including WB-PESQ \cite{941023}, CSIG, CBAK, and COVL \cite{4389058}. The results are presented in Table \ref{table1}. 

Based on DCTCRN, we respectively introduce the VAD aided multi-task learning scheme and the CSA block, so as to prove the effectiveness of the two ideas.

\textbf{VAD aided multi-task learning} Different from previous SE models, we have designed a dual-branch network for multi-task learning, where one branch is responsible for the SE task and the other for the VAD task. Indeed, the VAD task and the SE task exhibit a strong correlation, where the precise localization of speech segments by the VAD module proves helpful in enhancing the SE module's ability to suppress noise segments effectively and improve the quality of speech components. Specifically, the VAD aided SE model tends to preserve the original speech components when reducing the noise of the speech segments. As for the non-speech segments, it shows a stronger noise suppression capability to completely reduce the noise components. In Table \ref{table1}, we denote the VAD aided multi-task learning scheme as DCTCRN+VAD, and the corresponding results confirm its effectiveness. With the assistance of VAD, the WB-PESQ, CSIG, CBAK, and COVL of our DCTCRN increase by 0.07, 0.04, 0.04, and 0.03, respectively.

\textbf{CSA Block} In Table \ref{table1}, we introduce the CSA block to DCTCRN, denoted as DCTCRN+CSA. The CSA block aids the network in identifying important positions within the feature map, enabling it to analyze the key feature information. As a result, the experimental results demonstrate that the CSA block leads to performance improvements of 0.06 on WB-PESQ, 0.02 on CSIG, 0.04 on CBAK, and 0.03 on COVL.

Our VSANet combines the above two improvements, and its evaluation results are also presented in Table \ref{table1}. It can be found that by integrating the two proposed methods, the SE performance of our model is further significantly improved. Specifically, the two contributions result in a total improvement of 0.14 on WB-PESQ, 0.07 on CSIG, 0.06 on CBAK, and 0.08 on COVL.

\subsection{Comparison with the State-of-the-Art Methods}
We also compare our VSANet with other state-of-the-art (SOTA) methods, and the results are shown in Table \ref{tabel2}. To objectively demonstrate the performance benefits of our VSANet, all the benchmark methods are also implemented causally. Meanwhile, these benchmarks cover both time domain and TF domain methods. The TF domain methods utilize various types of spectrum features as input, including STFT magnitude spectrum and STFT complex spectrum. It can be observed that with the same CRN structure and smaller model size, the performance of our DCTCRN (as shown in Table \ref{table1}) is better than that of DCCRN \cite{hu20g_interspeech}, which illustrates the advantage of STDCT over STFT. 

Overall, our VSANet achieves the best scores on WB-PESQ, CBAK, and COVL, demonstrating its superior performance. Although the CSIG score of our VSANet is slightly lower than that of DEMUCS \cite{defossez20_interspeech} and GaGNet \cite{LI2022108499}, it is essential to note that our VSANet has a distinct advantage in terms of parameter size, with only 3.1M parameters, which is smaller than many existing methods. In addition, since our VSANet is causal, the algorithm delay is just one frame length, i.e., 32ms, which meets the real-time processing requirement \cite{reddy20_interspeech}. We have also deployed our VSANet on Intel(R) Xeon(R) Platinum 8255C CPU@2.50GHz, and the real-time factor is 0.15.

\section{Conclusion}
\label{sec:conclusion}
In this paper, we present a VAD aided multi-task learning framework and a CSA block to improve the performance of the STDCT-based SE method. We design a dual-branch network to simultaneously handle the SE and VAD tasks for noisy speech. The SE and VAD modules share the same encoder, but they have different recurrent blocks and decoders. During training, we optimize the SE and VAD modules jointly, leading to improved SE performance compared to ordinary single-task learning. In addition, we introduce a CSA block, which helps the network focus on important feature information, thereby enhancing its feature extraction and analysis capabilities. Meanwhile, the CSA block is designed to be causal, ensuring real-time SE. The ablation study confirms the effectiveness of the CSA block. Finally, we combine the above two innovations to propose the VSANet, which achieves superior real-time SE performance with a small model size. In the future, we plan to explore even more effective multi-task learning schemes, and we also plan to expand our methods to other related works, such as speech dereverberation and separation.

\section{Acknowledge}
\label{sec:acknowledge}
This work was supported by the special funds of Shenzhen Science and Technology Innovation Commission under Grant No. CJGJZD20220517141400002.

\clearpage
\bibliographystyle{IEEEbib}
\bibliography{refs}

\end{document}